\documentclass[useAMS,usenatbib]{mn2e}
\usepackage {graphicx}
\usepackage{natbib}
\usepackage{aas_journals}
\usepackage{color}
\usepackage{amsmath}
\newcommand{\mpc}{\rm {h^{-1}Mpc }}

\newcommand{\ltsima}{$\; \buildrel < \over \sim \;$}
\newcommand{\lsim}{\lower.5ex\hbox{\ltsima}}
\newcommand{\gtsima}{$\; \buildrel > \over \sim \;$}
\newcommand{\gsim}{\lower.5ex\hbox{\gtsima}}

\def\gtrsim{\mathrel{\hbox{\rlap{\hbox{\lower4pt\hbox{$\sim$}}}\hbox{$>$}}}}
\let\ga=\gtrsim
\def\lesssim{\mathrel{\hbox{\rlap{\hbox{\lower4pt\hbox{$\sim$}}}\hbox{$<$}}}}
\let\la=\lesssim

\newcommand{\muK}{\mu{\rm K}}

\title[Cold imprint of supervoids with Planck and BOSS DR10]{Cold imprint of supervoids in the Cosmic Microwave Background re-considered with Planck and BOSS DR10}
\author[Andr\'as Kov\'acs and Benjamin R. Granett]{Andr\'as Kov\'acs$^{1}$, Benjamin R. Granett$^{2}$\\
$^{1}$ Institut de F\'isica d'Altes Energies, Universitat Aut\'onoma de Barcelona, E-08193 Bellaterra (Barcelona), Spain\\
$^{2}$ INAF OA Brera, Via E. Bianchi 46, Merate, Italy}
\begin{document}

\date{Submitted 2015}

\pagerange{\pageref{firstpage}--\pageref{lastpage}} \pubyear{2014}

\maketitle

\label{firstpage}
\begin{abstract}

We analyze publicly available void catalogs of the Baryon Oscillation Spectroscopic Survey Data Release 10 at redshifts $0.4<z<0.7$. The first goal of this paper is to extend the Cosmic Microwave Background stacking analysis of previous spectroscopic void samples at $z<0.4$. In addition, the DR10 void catalog provides the first chance to spectroscopically probe the volume of the Granett et al. (2008) supervoid catalog that constitutes the only set of voids which has shown a significant detection of a cross-correlation signal between void locations and average CMB chill. We found that the positions of voids identified in the spectroscopic DR10 galaxy catalog typically do not coincide with the locations of the Granett et al. supervoids in the overlapping volume, in spite of the presence of large underdense regions of high void-density in DR10. This failure to locate the same structures with spectroscopic redshifts may arise due to systematic differences in the properties of voids detected in photometric and spectroscopic samples. In the stacking measurement, we first find a $\Delta T = - 11.5 \pm 3.7~\muK$ imprint for 35 of the 50 Granett et al. supervoids available in the DR10 volume. For the DR10 void catalog, lacking a prior on the number of voids to be considered in the stacking analysis, we find that the correlation measurement is fully consistent with no correlation. However, the measurement peaks with amplitude $\Delta T = - 9.8 \pm 4.8~\muK$ for the a posteriori-selected 44 largest voids of size $R>65~\mpc$ that does match in terms of amplitude and number of structures the Granett et al. observation, although at different void positions.
\end{abstract}
\begin{keywords}
surveys -- cosmology: observations -- large-scale structure of Universe -- cosmic background radiation
\end{keywords}

\section{Introduction}
Large-scale structures at low redshift leave their mark on the Cosmic Microwave Background (CMB) radiation providing direct probes of the late time cosmic acceleration and the physics of Dark Energy \citep{Aghanim2008}.  In particular, large voids and clusters can imprint themselves to the primary fluctuations of the CMB via physical mechanisms called the Integrated Sachs-Wolfe effect \citep[ISW]{SachsWolfe} in the linear regime, and the Rees-Sciama effect \citep[RS]{ReesSciama} on non-linear scales. The expected ISW correlation in the $\Lambda CDM$ model is on the order of $0.1 ~\muK < |\Delta T | < 1 ~\muK$ for typical voids \citep{CaiEtAl2010}, extending up to $|\Delta T | \approx 20 ~\muK$ for the largest observable superstructures which are also the rarest (see e.g. \cite{SzapudiEtAl2014,Nadathur2014}). The contribution of the non-linear RS effects remains typically at the $\sim 10 \%$ level compared to the linear expectation \citep{CaiEtAl2010}. However, the ISW and RS effects and their relative strength may be different in alternative cosmologies \citep{CaiEtAl2014}.

The typical ISW and RS imprints are thus small enough to be immeasurable in practice. The traditional approach for detecting the weak signal is the angular cross-correlation measurement between galaxy density maps and the CMB. This detection strategy has been followed by a series of studies finding both marginally (see e.g. \cite{b14}, \cite{KovacsEtAl2013}) and moderately significant (see e.g. \cite{ho,gianEtAl08,gian,Planck19} and references therein) ISW-like signals.
Another approach is focussed on the largest structures in the density field, where the ISW-RS effect is expected to be the strongest. Foremost, \cite{GranettEtAl2008} created a catalog of supervoids and superclusters\footnote{\texttt{http://ifa.hawaii.edu/cosmowave/supervoids/}} (Gr08, hereafter) using the SDSS Data Release 4 (DR4) Mega-z photometric LRG catalog \citep{megaz} with some additional area from DR6. They used the \texttt{ZOBOV} algorithm\footnote{\texttt{http://skysrv.pha.jhu.edu/neyrinck/voboz/}} \citep{ZOBOV} to identify the most prominent extremes of the large-scale density field. The superstructures were then used for stacking the CMB temperature centered on these locations, and to measure an average effect through a compensated top-hat filter. 

This rather simple statistic averages the CMB temperature centered on voids within a circular aperture $r < R$, and then subtracts the background temperature averaged in an equal-area concentric annulus with $R < r < \sqrt 2R$. \cite{GranettEtAl2008} found a $|\Delta T| = 9.6 \pm 2.2 ~\muK$ signal for their 100 most significant ($>3\sigma$) superstructures using an aperture size of $R=4^{\circ}$. This signal appears to be in $\gsim2 \sigma$ tension with $\Lambda CDM$ predictions, as pointed out in several follow-up studies using theory and simulations \citep{PapaiEtAl2011,PapaiSzapudi2010,Nadathur2012,Flender2013,Hernandez2013,Hotchkiss2015,Aiola}. The expectation for the stacked ISW signal remains at the $|\Delta T| \lsim 2 ~\muK$ level for $\sim 50$ superstructures similar to Gr08 objects.

Also, numerous additional tests have been performed to uncover possible systematic problems and statistical biases \citep{Ilic2013,Planck19,CaiEtAl2013}. It was found that varying the number of the objects in the stacking, or using different filter sizes typically lowers the overall significance. Otherwise the original Gr08 signal has survived every revision and remains a puzzle.

Additionally, \cite{Planck19}, \cite{CaiEtAl2013}, and \cite{Ilic2013} repeated the CMB stacking analysis of \cite{GranettEtAl2008} using complementary void catalogs \citep{Sutter2012} based on spectroscopic measurements at $z<0.4$. These studies, however, did not report a highly significant detection of the ISW-like effect found in Gr08, except the $\sim 2\sigma$ evidence for a correlation in \cite{CaiEtAl2013}.

Due to the overwhelming contribution from cosmic variance, ISW stacking measurements can be prone to misinterpretation due to the `look elsewhere' effect \citep{Peiris2014}.  The measurement parameter space has a high dimensionality when counting the choices made in the void catalogue selection and methodology details such as filter size, and the inherently weak signal can lead to the over interpretation of statistical flukes \citep{Hernandez2013}.  

In this work we do not formally carry out a blind analysis to mitigate the effects; however, we fix the measurement methodology using parameters determined externally.  We test the robustness of the signal by varying these parameters, expecting that a true signal will be robust to perturbations in the catalogue properties or methodology. We compare the Gr08 void catalog to the Baryon Oscillation Spectroscopic Survey (BOSS) DR10 CMASS and LOW-Z void catalogs provided by \cite{Sutter_DR10}. On one hand, we are able to probe a significant fraction of the Gr08 volume (surveyed with photo-z data) now with  voids detected using spectroscopic redshifts. On the other hand, we extend previous low-z DR7 void stacking measurements to the range of $0.4<z<0.7$. Two distinct conclusions are possible: our stacking analysis could confirm the \cite{GranettEtAl2008} detection for the first time with an independent void catalog, or the ISW(-like) signal could disappear. In any case, some puzzle will certainly remain for future analyses with BOSS DR12 and {\it Planck} DR2.

The paper is organized as follows. Data sets, algorithms, and our observational results are presented in Section 2; the final section contains a summary, discussion and interpretation of our results. 

\section{Data sets and measurements}
\begin{figure*}
\begin{center}
\includegraphics[width=160mm]{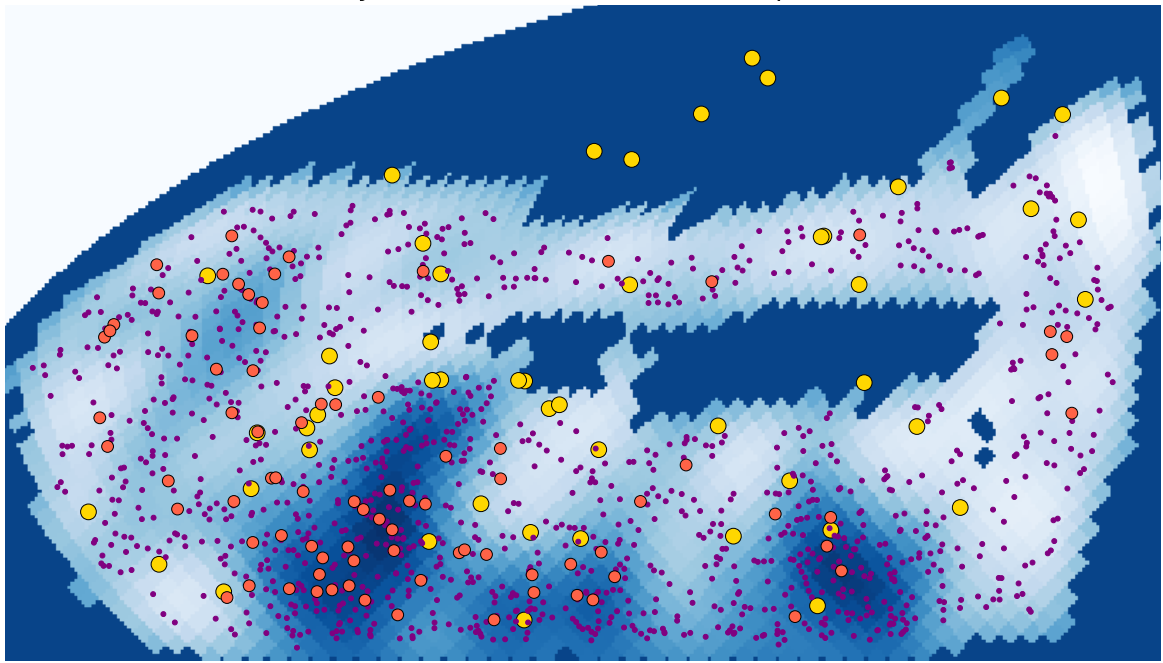}
\label{fig1}
\caption{Void positions of the Gr08 sample (gold) vs. DR10 void catalog (purple). Red points show a subclass of large DR10 voids of size $R_{v} > 60~\mpc$. The underlying $N_{side}=32$ \texttt{HEALPix} map is the DR10 void density sample smoothed with a $\sigma =4^{\circ}$ Gaussian. Due to the hierarchical organisation of the void catalogue, an underdense region may be split into a number of voids of various sizes in the catalogue.  Dark blue colors indicate higher void abundance thus lower average projected density. The fact that the two void catalogues trace different structures points to systematic differences in the galaxy catalogues and void finding algorithms.}
\end{center}
\end{figure*}

\subsection{CMB data}
On the CMB side we use {\it Planck}'s SMICA\footnote{\texttt{http://www.cosmos.esa.int/web/planck}} map \citep{Planck_15} with resolution downgraded to $N_{side}=512$ with the \texttt{HEALPix}\citep{healpix} pixelization. We mask out potentially contaminated CMB pixels using the WMAP 9-year extended temperature analysis mask\footnote{\texttt{http://lambda.gsfc.nasa.gov/product/map/dr5/}} \citep{WMAP9} at $N_{side}=512$ to avoid repixelization effects of the $N_{side}=2048$ CMB masks provided by {\it Planck}. It has already been pointed out by \cite{GranettEtAl2008}, and later confirmed by \cite{Ilic2013}, \cite{Planck19}, and \cite{CaiEtAl2013} that the ISW-like cross-correlation signal detected at void locations is independent of the CMB data set when looking at WMAP Q, V, W, or {\it Planck} SMICA temperature maps. We thus limit our analysis to the latest {\it Planck} SMICA sky map.
\subsection{Catalogs of cosmic voids}

We use public void catalogs by \cite{Sutter_DR10} where the authors identified voids in BOSS DR10 spectroscopic galaxy samples \citep{BOSS_DR10}. The voids were identified with the \texttt{ZOBOV} tool \citep{ZOBOV} within the \texttt{VIDE}\footnote{\texttt{http://www.cosmicvoids.net}} framework \citep{VIDE}. The void-finder \texttt{ZOBOV}  is based on the watershed algorithm which builds a hierarchy of underdensities.  We first remove voids with size $R<40~\mpc$ to cut the hierarchy.

Supervoids encompassing smaller voids, walls, and filaments may be represented by multiple voids in a dense tracer catalog, or by one large void in sparsely sampled data or in the presence of photometric redshift uncertainties \citep{Sutter_DM}. The galaxy number density for the CMASS catalog peaks at z=0.5 with mean inter-galaxy separation $L \approx 16~\mpc$ which rises to $\sim25~\mpc$ at $z=0.65$, thus a lower cut of roughly twice this characteristic scale is a safe and reasonable choice to prune spurious void detections that would contribute only noise to the measurement \citep{Sutter_DR10,VIDE}.

Furthermore, these small and potentially spurious voids may occupy over dense regions. \cite{CaiEtAl2013} tested this effect in mock catalogs.  They suggest a lower size cut of $R>65~\mpc$ for a complete removal of potentially spurious voids at $z_{med}\approx0.43$ in their {\it lrgbright} sample ($L \approx 38~\mpc$), i.e. the DR7 subsample that is possibly the most similar to the better sampled DR10 data we consider here. Although it is not possible to adopt such cuts for DR10 voids without proper simulations of the source catalogs, a rough comparison of the source densities of DR7 and DR10 catalogs indicates that $R > 40~ \mpc$ might be a reasonable cut.

We also restrict our analysis to central voids to minimise possible contaminations caused by the survey mask (see e.g. \cite{Sutter_DR10} for details).  

These moderately conservative cuts remove $\sim 65 \%$ of the voids from the DR10 catalog.  Our approach is to probe the $0.4<z<0.7$ redshift range thus we  consider CMASS data in our analysis, with 13 extra voids from the LOW-Z 4 sample at $0.4<z<0.45$, i.e. a redshift range of the Gr08 catalog not covered by CMASS data.

Following \cite{Sutter_DR10}, we divide the resulting DR10 void catalog into three subsamples; a combined CMASS 1 + LOW-Z 4 at $0.4<z<0.5$ (56+13 voids), CMASS 2 at $0.5<z<0.6$ (237 voids), and CMASS 3 at $0.6<z<0.7$ (172 voids). We analyze these catalogs both separately and jointly.
We removed 6 voids from the analysis, as their position was curiously outside of the rough $N_{side}=32$ DR10 footprint by $\sim 2^{\circ}$, indicating an inconsistency in the void catalog\footnote{We consulted Sutter et al. about this issue, and learned that these objects will be revised in a later version of their DR10 void catalog, and should be removed at the moment.}. We checked the effects of these objects on our main results and found no difference. 

We show our void sample together with the 50 Gr08 supervoids in Figure 1 (35 of the 50 Gr08 voids should be detectable in DR10, i.e. not masked out or residing close to the boundary). On average DR10 voids are smaller in angular and physical size than the Gr08 superstructures, due to the ability of resolving small-scale structures with spectroscopic redshifts. However, we found large regions of high density of DR10 voids, which typically do not overlap with the larger Gr08 supervoids. Thus the expected fragmentation of superstructures into smaller voids is not observed, but additional large underdense regions appear in the spectroscopic data. This somewhat counter-intuitive finding means that a potential (and expected) ISW(-like) signal in the shared DR10-Gr08 volume is in this case carried by voids at distinct locations in the sky.   

The number densities of the two catalogues are of similar order: while the CMASS tracer inter-galaxy separation is $\bar{n}^{-1/3} \approx 14-22~\mpc$, the photometric LRG sample is $\sim2.2$ times more dense  with $\bar{n}^{-1/3}\approx 10-17~\mpc$. 

The largest DR10 voids (minimum as large as the typical Gr08 supervoids), however, are located closer to the Gr08 supervoids on average. DR10 voids of size $R>60~\mpc$ are shown separately in Figure 1. Appropriate analysis of the properties of this subclass of voids is one of the main goals of this paper.

\subsection{Methods \& Results}

\begin{figure*}
\begin{center}
\includegraphics[width=190mm]{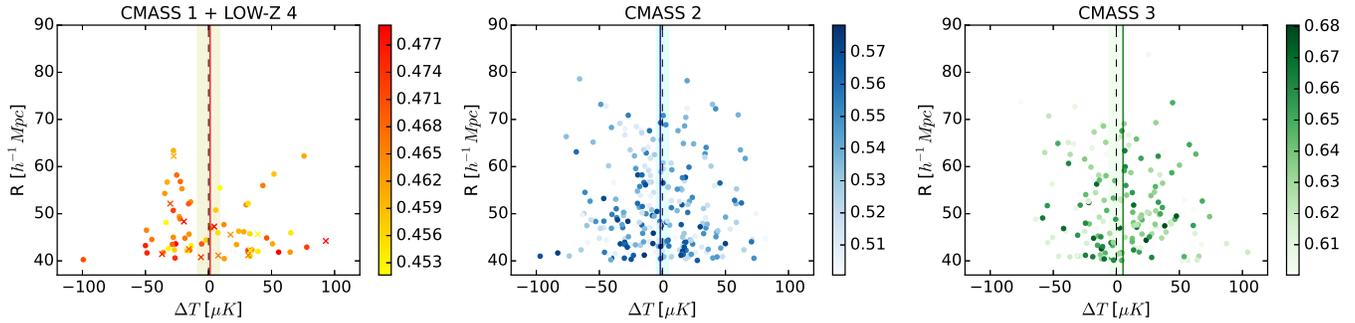}
\label{T_rad}
\caption{Filtered temperatures in re-scaled top-hats are shown as a function of their physical size. Crosses in the left panel mark LOW-Z 4 voids. Color bars indicate the redshifts of the voids, without any apparent trend or clustering in this parameter space. We note, however, that there is a slight extra average cooling for small CMASS 2 voids. Interestingly, CMASS 3 voids behave inversely showing hotter temperature differences on average for the smallest voids. Shaded regions mark out $2\sigma$ statistical uncertainties scaled with the number of objects considered, while solid lines indicate the stacked temperature for a given subsample.}
\end{center}
\end{figure*}

Foremost, we repeated the stacking analysis of \cite{GranettEtAl2008} for the 35 supervoids available in the DR10 volume with constant $R=4^{\circ}$ filter radius. The original signal of $\Delta T = - 11.3 \pm 3.1~ \muK$ signal as measured by \cite{GranettEtAl2008} now limited to the DR10 area is changed to $\Delta T = - 11.5 \pm 3.7~ \muK$. We then expected to detect a similar signal in the same physical volume with voids identified using spectroscopic redshift from the DR10 CMASS catalog.

In our methods, we closely follow \cite{Ilic2013} and \cite{CaiEtAl2013}. We first measure average temperatures in the SMICA map at void locations using the compensated top-hat filter technique applied by \cite{GranettEtAl2008}.  We further scale the filter by angular size as advanced by \cite{Ilic2013}, \cite{CaiEtAl2013}, and \cite{Hotchkiss2015}. The same authors empirically found in data and in simulations that the optimal filter size to maximize $S/N$ is $\sim 60 \%$ of the re-scaled void radius. The physical motivation behind such a scaling is the coincidence with the zero crossings of the void density profile and the cumulative ISW-RS signal found in N-body simulations \citep{CaiEtAl2013}.
We adopt this refinement in order to maximize the expected signal-to-noise in our tests. The resulting typical re-scaled filter radius is $r_{mean}\approx1.2^{\circ}$ for all sub-samples. 

Our findings are presented in Figures 2 and 3, showing the void temperature measured as a function of void radius and redshift. We compare the redshift distributions of DR10 and Gr08 voids in Figure 3. These plots show the typical behavior of such top-hat filtered temperatures, as they contain large fluctuations for individual objects of both positive and negative signs. However, there is no obvious excess clustering or other oddity in these parameter spaces. Two exceptions are the slight average shift to the negative side for $R < 50~\mpc$ CMASS 2 objects, and the counteractive change at $R < 50~\mpc$ for CMASS 3. Note that these voids should carry the lowest ISW-RS signal among the catalog, and their robustness is questionable due to the occurrence of voids in clouds \citep{CaiEtAl2013}.

\begin{figure}
\begin{center}
\includegraphics[width=80mm]{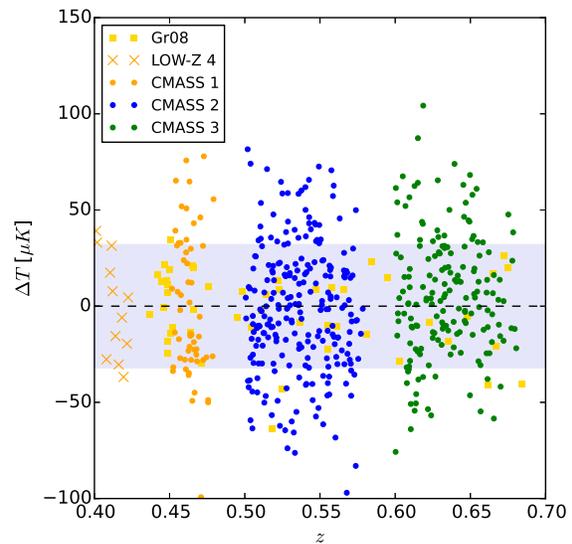}
\label{T_reds}
\caption{Filtered temperatures in re-scaled top-hats are shown as a function of their redshift, covering the full range of Gr08 voids. No meaningful trend is observable, as all sub-samples show similar distributions. The grey shaded region marks the $1\sigma$ fluctuation $\sigma_{\Delta T} \approx 32~ \muK$ for a single object, as measured using CMB simulations.}
\end{center}
\end{figure}

\begin{figure}
\begin{center}
\includegraphics[width=90mm]{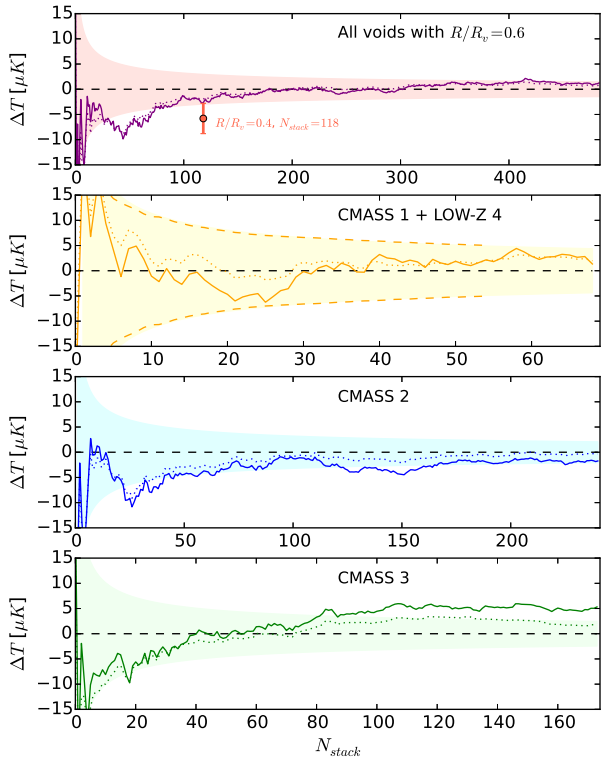}
\caption{Stacked CMB temperatures as a function of the number of the objects considered. A physically motivated ordering of the voids by radius is applied, as the largest voids should leave the coldest imprints on the CMB. Dotted lines represent our results when an additional weighting by void probability is applied. Orange dashes in panel 2 mark the errors obtained by measuring standard deviations for all filter sizes in simulations independently. In the top panel, we compare our results to the best signal-to-noise achieved by non-optimal rescaling strategy. \label{stackings}}
\end{center}
\end{figure}

We then average the filtered temperatures for the 478 DR10 voids, sorted by radius. We estimated statistical uncertainties by analyzing $1000$ Gaussian CMB simulations generated with the \texttt{HEALPix} \citep{healpix} \texttt{synfast} routine using the {\it Planck} DR1 best fit CMB power spectrum \citep{Planck_15}.  Gaussian simulations without considering instrument noise suffice because the CMB signal is dominated by cosmic variance on the scales we consider (See e.g. \cite{Hotchkiss2015}). 

The ISW-RS signal expected in $\Lambda CDM$ is so small that it is dominated by the uncertainty of the primary anisotropies even with stacking applied. More precisely, $S/N \lsim 0.4$ for filters $R<2^{\circ}$ was estimated by \cite{Flender2013}. \cite{Nadathur2012} pointed out that $N_{stack}\approx 3000$ supervoids could provide $S/N \sim 2.5$ for $\Delta T \approx - 2 ~\muK$, i.e. a signal that can be produced by the most extreme superstructures in $\Lambda CDM$. Unfortunately, current observational capabilities cannot provide such a numerous catalog of voids.

We adapt the error analysis for the stacking presented e.g. in \cite{GranettEtAl2008}, \cite{Flender2013}, and \cite{CaiEtAl2013}, and compare two error estimators.  
First, we repeated the stacking analysis $1000$ times varying the CMB realization and fixing the position and scaling of the 56 top-hat filters from the CMASS 1 data. We compare this against a simpler approach in which the variance is computed for a single filter (assumed to be independent) and rescaled by the number of voids in the catalogue. The mean variance is estimated from 1000 realizations but on each realization the filter size is randomly drawn from the size distribution. This gives a mean standard deviation of  $\sigma_{\Delta T} \approx 32~ \muK$ and we find that this error does not depend on the redshift bin.  We then scale the variance by the number of objects in the stack: 
$\sigma_{\Delta T}(N_{stack}) = 32~ \muK / \sqrt{N_{stack}}$. This approach simplifies the computation and tests revealed that the two estimators agree at the per cent level based on the standard deviation measurement. The CMASS 1 panel of Figure 4 contains a comparison of the error bars obtained with the two methods showing good agreement. 
This finding is in agreement with that of \cite{GranettEtAl2008} confirming that the measurements with the compensated filter may be considered to be independent. \cite{GranettEtAl2008} also pointed out that errors obtained by drawing random points to a given CMB map, and errors measured with fixed void positions varying the CMB realization agree at the $\sim2\%$ level.

The mean error we obtained is higher than the $\sigma_{\Delta T} \approx 22~ \muK$ uncertainty for a single supervoid found by \cite{GranettEtAl2008}, due to larger CMB fluctuations at smaller scales. Note that this higher noise level prevents high signal-to-noise measurements using relatively small voids.

This procedure allows us to estimate the significance of a measurement given a fixed number of voids $N_{stack}$.  However, with uncertainties in the properties of the voids and the origin of the signal we do not have a strong prior on the number to average in the analysis.  Using only the largest voids we do not have the precision to measure a weak signal,  while adding the smaller voids we can wash-out a signal if it exists.

\begin{figure}
\begin{center}
\includegraphics[width=90mm]{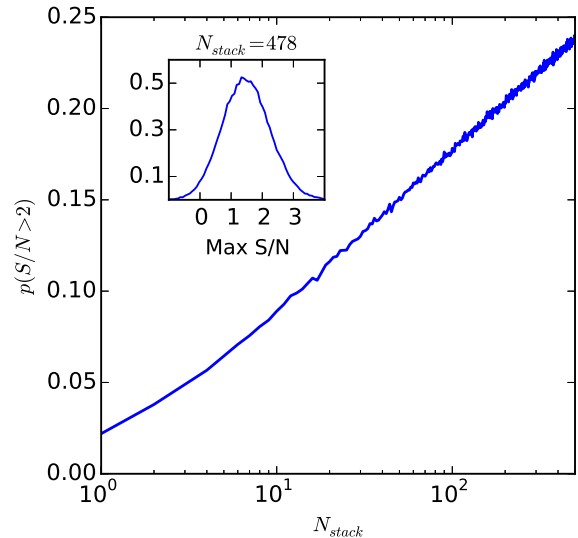}
\caption{The probability of finding at least one spurious signal at the negative extreme with $S/N>2$ is plotted as a function of $N_{stack}$ in the Gaussian noise model. The inset shows the probability distribution of maximum $S/N$ for the case of all voids considered in the stacking.\label{errorsrev}}
\end{center}
\end{figure}

To address this situation, \cite{GranettEtAl2008} define a cutoff based upon void probability, taking all voids with detection signficance $>3\sigma$.  Further, \cite{CaiEtAl2013}  applied a weighting as a function of void probability; however, they found no significant difference in the measured signal.  We also test this weighting scheme here, as shown in Fig. 4 and discussed below.

Without using a prior as to which voids to consider in the measurement we are led to examine the signal as a function of $N_{stack}$ ordered by void size, as in Fig. \ref{stackings}.  The temperature measured after averaging $N_{stack}$ structures appears as a random walk and, given the null hypothesis of no correlation, we may estimate the probability that the trajectory crosses a given threshold for detection.  We compute this probability in Monte Carlo fashion assuming the Gaussian noise model.  We generate a vector of $N_{tot}$ Gaussian distributed values with $\sigma=32~\muK$.  We then check if the cumulative sum crosses a given significance threshold at any point.   This is repeated and we keep count of the fraction of runs that give a significance above 2$\sigma$.  The result after 100000 runs is shown as a function of $N_{stack}$ in Fig. \ref{errorsrev}.  We find that given a catalogue of 478 voids, the odds of finding a 2$\sigma$ signal in the cumulative temperature measurement given the null hypothesis is 24\%.

We now consider the stacked temperature measurements shown in Fig. 4.  
\begin{enumerate}
\item Considering the combined sample (Fig. 4., top panel), the signal fluctuates around the 1$\sigma$ level.  A cold imprint peaks with an amplitude of $-9.8 \pm 4.8 ~\muK$ or $\sim 2\sigma$ for the 44 largest voids with sizes $R > 65~\mpc$.  However, this is unremarkable given that the probability of finding such an extreme somewhere in the cumulative stacked temperature with a total of 478 structures is $\sim24$\%.
\item However, the Gr08 sample contains 50 significant supervoids, thus in terms of the largest $\sim$50 fluctuations the two catalogs seem to agree with each other even if the structures differ in size and are not at the same positions.
\item Counting all 478 voids down to $R=40 ~\mpc$ the signal sharply approaches to zero and becomes $\Delta T \approx 1.2 \pm 1.5 ~\muK$. Note that the overall measurement, including the peak, is unchanged when weighting by void probability.
\item Applying a presumably non-optimal filter scaling of $40 \%$ of the void radius, we found $-5.8 \pm 3.0 ~\muK$ or $\sim 1.9\sigma$ for the 118 largest voids with sizes $R > 55~\mpc$. This signal, shown by the error bar in the top panel of Figure 4, is the largest  measured using different rescaled filter sizes. Note that this signal is also observed at relatively large void radii, and it is consistent with our measurement at $N_{stack}=118$ using $R/R_{v}=0.6$.
\item The combined CMASS 1+LOW-Z 4 sample shows a signal fluctuating inside the $1\sigma$ level, resulting in a final value of $\Delta T \approx 1.3 \pm 3.9 ~\muK$ for the full sub-catalog with 69 void members. Note the effect of the probability weighting which reduces the amplitude.
\item The CMASS 2 sample with the largest number of voids contributes most strongly to the combined sample, thus the signal is similar to that described above. After fluctuating inside the $1\sigma$ expectation and reaching the $\sim 1.5\sigma$ level or $\Delta T \approx -10~\muK$ at $R > 65~\mpc$, the overall signal-to-noise with all voids included is $\Delta T \approx -1.7 \pm 2.1 ~\muK$ for the 237 CMASS 2 objects. The weighting by void probability results in a more effective convergence to zero.
\item The CMASS 3 temperature signal fluctuates around the 1$\sigma$ level.  Adding smaller scale voids the signal becomes positive and results in a curious positive $\Delta T \approx 5.3 \pm 2.5 ~\muK$. However, the galaxy number density is lowest in the CMASS 3 bin, so a larger fraction of the smallest voids may be spurious compared with the other more densely sampled redshift bins. This intuition is verified by the probability weighting test, that lowers the significance of the spurious positive signal, resulting in  $\Delta T \approx 2.0 \pm 2.5 ~\muK$.
\end{enumerate}

\begin{figure}
\begin{center}
\includegraphics[width=90mm]{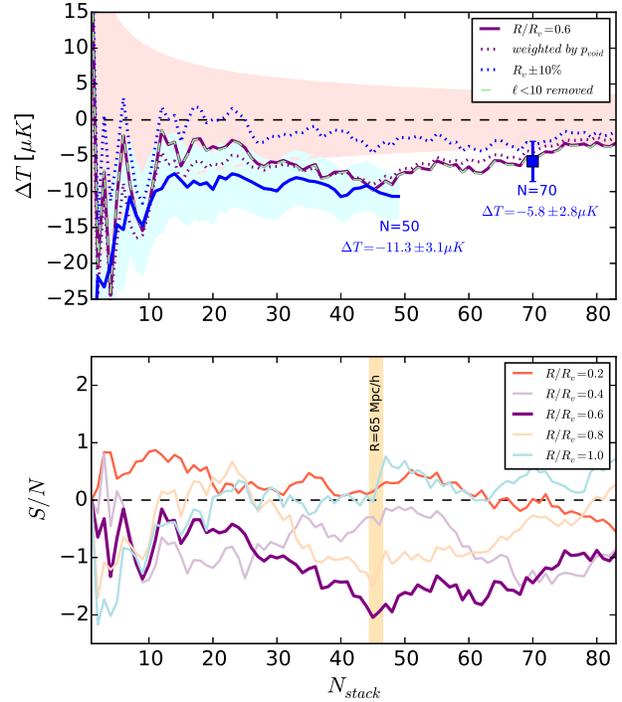}
\label{stackings_large}
\caption{Stacked CMB temperatures for voids of size $R_{v} > 60~\mpc$ ordered by radius. In the bottom panel, we show the signal-to-noise ratio of our measurement for different re-scaling parameters. In the top panel, we compare various measurement strategies for case $R/R_{v}=0.6$, i.e. the best filter size obtained in both simulation and data. The maximal signal-to-noise is observed at the lower size limit of $R_{v} > 65~\mpc$. The signal we detected is comparable to the Gr08 observation (shown in blue), although less significant.}
\end{center}
\end{figure}

\begin{figure}
\begin{center}
\includegraphics[width=90mm]{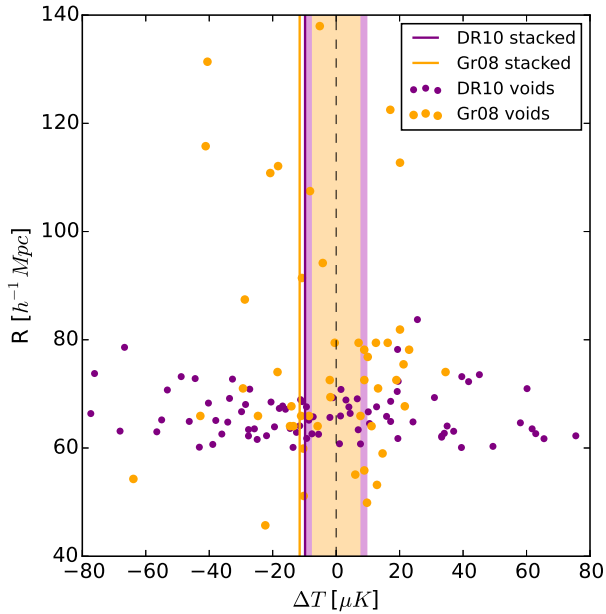}
\label{stackings_large_comp}
\caption{A comparison of filtered CMB temperatures as a function of the physical size of voids. Shaded regions mark $2\sigma$ fluctuations, while solid lines show the measurements by Gr08 and a result obtained using our final sub-catalog.}
\end{center}
\end{figure}

\subsection{Imprint of DR10 supervoids}

We have an indication that the largest structures in DR10 may produce an imprint on the CMB characterised by the $-9.8 \pm 4.8 ~\muK$ or $\sim 2\sigma$ signal that we measured for voids of size $R > 65~\mpc$.  Although on its own the occurrence of the signal is unremarkable in the light of our Monte Carlo tests, it does match the Gr08 measurement in terms of amplitude and number of structures.  Thus this result warrants further inspection. Our findings, based on additional systematic tests are shown in Figure 6, and summarized below
\begin{enumerate}
\item the lower panel of Figure 6 demonstrates that the $R/R_{v}=0.6$ scaling gives the highest $S/N$, as found in simulations and DR7 data.
\item the shape of the dotted purple curve in the upper panel indicates that this signal is robust against weighting by void probability, in agreement with \cite{CaiEtAl2013}.
\item we randomly changed the void radius by $+10\%$ or $-10\%$ for all voids, and the signal nearly disappeared, as shown in blue.
\item we artificially removed large-scale fluctuations (i.e. much larger than the void size) from the SMICA CMB map using \texttt{HEALPix} routines, and repeated the stacking. Without $\ell<10$ modes, the map has changed, and errors decreased by $\approx3\%$, but the filtered temperatures remain almost the same (shown with green dashes). This finding is again consistent with the DR7 analysis by \cite{CaiEtAl2013}.
\item despite the different stacking methods and void properties, the signal we found is in $1\sigma$ agreement with Gr08, although less significant due to larger CMB fluctuations at smaller angular scales (Gr08: constant $R=4^{\circ}$ filter, DR10: $R/R_{v}=0.6$ re-scaling)
\end{enumerate}

Therefore, we see that the our measurement appears to be robust against physically motivated changes in the analysis, while the weak signal disappears when non-optimal techniques are applied.
The exception is when the void radii were perturbed by 10\% and we find a reduction in the signal.  An inherent uncertainty is expected in the void radii due to sampling error and variation in void shape, so this may point to a fragility of the signal.  

Keeping in mind that the ISW-RS signal expected in $\Lambda CDM$ remains below $\Delta T \approx -2.0 ~\muK$ even for the 50 largest posteriori-selected supervoids we also lack the theoretical motivation for this signal. 

While the $4.4\sigma$ signal for the posteriori-selected, 50 most significant Gr08 superstructures appears to be too large for a plausible primordial fluctuation, \cite{PlanckISW2015} explicitly pointed out that the lack of a strong $T\sim E$ correlation in the CMB polarization data by {\it Planck} in the location of the Gr08 structures provides evidence that the effect is not a primordial fluctuation.
Also, systematic effects are obvious suspects, but explanations such as spurious correlations caused by stellar contamination, extragalactic radio sources, or contamination via the Sunyaev-ZelÕdovich (SZ) effect have already been excluded by Gr08 with conservative masking and probes of CMB color dependence further constrained by Planck analyses  \citep{PlanckISW2015}.
Another uncertainty in the superstructure catalog, introduced by $\sigma_{z}\approx 0.05$ photometric redshift errors of Mega-z LRGs, have not been investigated. 

Figure 7 shows a comparison of filtered temperatures for Gr08 and DR10 voids as a function of their radius. In summary, the main properties are as follows.
\begin{enumerate}
\item larger fluctuations are observed in DR10 results due to smaller filter size.
\item proportionally more $R > 65~\mpc$ (super)voids were found in Gr08 than in DR10, possibly due to systematic differences in void detection and source catalogs.
\item in Gr08, the largest supervoids leave an ever stronger stacked imprint in the CMB.
\end{enumerate}

Large underdense regions of diameter $\sim 130~\mpc$, perhaps better thought of as long wavelength fluctuations, can contain smaller voids, walls, and filaments. Therefore, to identify these supervoids in galaxy surveys it will be necessary to modify void detection algorithms. The voids effectively resolved by spectroscopic redshift surveys may be grouped together and unified to identify under-densities on the largest scales.  Focusing on the class of supervoids would further simplify the interpretation of results by limiting the selection effects arising from {\it a posteriori} choices and illuminate the possible connection between the apparent ISW excess and the density field on the largest scales.

\section{Conclusions}

We probed the volume of the \cite{GranettEtAl2008} supervoid catalog with the BOSS DR10 void catalog provided by \cite{Sutter_DR10}. Our principal aim was to revisit the strong ISW(-like) signal found in Gr08 with a catalog that probes the same density field. We pruned the DR10 catalog following the protocol of \cite{CaiEtAl2013} and the suggestion of \cite{Sutter_DR10}  for removing the smallest and least reliable voids which are also expected to produce the smallest ISW-RS signals in $\Lambda CDM$. 
The voids identified with spectroscopic redshifts in DR10 are smaller than the Gr08 structures traced with photometric redshifts.  Even so, we find that the Gr08 supervoid positions do not coincide with regions abundant in DR10 voids which indicates that the void finders are sensitive to different structures. This situation merits further study to understand the systematic differences between voids identified in spectroscopic versus photometric samples.

We measured the stacked CMB temperatures for the 35 Gr08 supervoids in the DR10 footprint using the original filter size. The $\Delta T = - 11.5 \pm 3.7 ~\muK$ signal that we found is consistent with the Gr08 measurement. We then performed a stacking analysis using 478 DR10 CMASS and LOW-Z voids. In general, no significant imprint was detected in the presence of large cosmic variance. However, we found a $\Delta T = - 9.8 \pm 4.8~ \muK$ or $2\sigma$ signal for the largest 44 voids of size $R > 65~\mpc$ in the combined sample. We note that this detection in itself is unremarkable without a strong prior on the number of sources or minimum void size used in the analysis, as we found that the probability of finding a 2$\sigma$ signal somewhere in the cumulative temperature measurement with 478 voids is $p\approx24\%$.  However, the measurement is of interest considering the detection based on the Gr08 sample.
We examined how robust the measurement is to variations in the methodology.  Varying the filter size, we found that the signal is maximised when using a filter radius found to be optimal in N-body simulations.

Our results highlight that ISW detections with the stacking protocol strongly depend on the properties of the tracer population and the void finder. While the effect of photo-z errors on the performance of \texttt{ZOBOV} have not been tested, it has been emphasized by \cite{Sutter_DR9} that masking and the density of the tracer population strongly affects the resulting void catalogs.  Further systematic uncertainties remain in how the hierarchy of voids is cut and the location of void centers.

\section*{Acknowledgments}
We thank Mark Neyrinck and Yanchuan Cai for giving comments that improved the paper.  AK takes immense pleasure in thanking the support provided by the Campus Hungary fellowship program. Funding for this project was partially provided by the Spanish Ministerio de Econom'a y Competitividad (MINECO) under projects FPA2012-39684, and Centro de Excelencia Severo Ochoa SEV-2012-0234.  BRG acknowledges support of the European Research Council through the Darklight ERC Advanced Research Grant (\# 291521).

Funding for SDSS-III has been provided by the Alfred P. Sloan Foundation, the Participating Institutions, the National Science Foundation, and the U.S. Department of Energy Office of Science. The SDSS-III web site is http://www.sdss3.org/.
SDSS-III is managed by the Astrophysical Research Consortium for the Participating Institutions of the SDSS-III Collaboration including the University of Arizona, the Brazilian Participation Group, Brookhaven National Laboratory, Carnegie Mellon University, University of Florida, the French Participation Group, the German Participation Group, Harvard University, the Instituto de Astrofisica de Canarias, the Michigan State/Notre Dame/JINA Participation Group, Johns Hopkins University, Lawrence Berkeley National Laboratory, Max Planck Institute for Astrophysics, Max Planck Institute for Extraterrestrial Physics, New Mexico State University, New York University, Ohio State University, Pennsylvania State University, University of Portsmouth, Princeton University, the Spanish Participation Group, University of Tokyo, University of Utah, Vanderbilt University, University of Virginia, University of Washington, and Yale University.

\bibliographystyle{mn2e}
\bibliography{refs}
\end{document}